\documentclass[aps,prl,reprint,showpacs,floatfix]{revtex4-2}
\usepackage{amsmath,amsfonts,amssymb,graphics,graphicx,epsfig,color,times,indentfirst,layout,hyperref,subfigure,multirow,bm}
\usepackage{hyperref, soul}
\usepackage{ulem}
\hypersetup{linktocpage,,colorlinks=true}
\usepackage{threeparttable}
\usepackage{dcolumn}
\usepackage{bm}
\usepackage{color}

\begin{document}
	
\title{Optical Signatures of Band Flatness and Anisotropic Quantum Geometry in Magic-Angle Twisted Bilayer Graphene}
\author{Pok Man Chiu}\email{pmchiu2022@gmail.com}
\affiliation{Graduate Institute of Applied Physics, National Chengchi University, Taipei City 11605, Taiwan}
\affiliation{Department of Physics, National Tsing Hua University, Hsinchu 30013, Taiwan}

\date{\today}

\begin{abstract}
We study the degree of band flatness and anisotropic quantum geometry in magic-angle twisted bilayer graphene by varying the twist angle and the lattice relaxation through optical conductivity. We show that the degree of band flatness and its quantum geometry can be revealed through optical absorption and its resulting optical bounds, which are based on the trace condition in quantum geometry. More specifically, the narrow and isolated peak of optical absorption in the low-energy region provides information about the bandwidth between two flat bands. When this value is smaller than the electron interaction, it serves as a critical condition for the emergence of flat band superconductivity. Furthermore, optical absorption also provides the gap value between the flat band and the dispersive band, and when this gap is larger than the electron interaction, it facilitates the realization of fractional Chern insulating phases. We show that the narrow and isolated peak of optical bound near zero energy decreases as lattice relaxation increases. Meanwhile, we demonstrate that the imaginary part of generalized optical Hall conductivity reveals the vanishing of the negative part of Berry curvature, which is enforced by the refined trace-determinant inequality. Accordingly, we show that the total amount of the negative part and component of the Berry curvature approaches zero in the single ideal flat-band case. In contrast, when considering all occupied bands, the total amount of the negative component is slightly different from zero. Finally, we demonstrate that the condition of vanishing of flat band velocities and the emergent chiral symmetry are sufficient for the saturation of the trace condition, which pertains to the isotropic case. In contrast, the determinant condition can be saturated in anisotropic systems, and the quantum metric and Berry curvature are related by a "saturation matrix" whose elements can be measured through the optical response. On the other hand, the non-vanishing difference between the optical bound and the imaginary part of the generalized optical Hall conductivity also provides an anisotropic signature in the fractional Chern insulating phases.
\end{abstract}

\maketitle


\textit{Introduction}.---Moiré superlattice materials provide ideal platforms for engineering strongly correlated phases of matter, in which flat bands are believed to play a key role. In these two-dimensional van der Waals heterostructures, the crucial tuning parameter is the twist angle, which leads to the emergence of nearly flat bands and the resulting strongly correlated phenomena. The most intriguing problem is the realization of flat band superconductivity and fractional Chern insulator (FCI) in moiré materials. In 2018, the first experiment \cite{Cao18} successfully demonstrated flat band superconductivity in twisted bilayer graphene (TBG) near the magic angle \cite{MacDonald11}. Recently, an experimental group discovered the FCI phase in TBG under a weak magnetic field (5 T) \cite{Xie21}. Most recently, several groups have experimentally observed the fractional quantum anomalous Hall effect in twisted bilayer $\text{MoTe}_2$ \cite{Park23,Cai23,Zeng23,Xu23,Anderson24,Redekop24}. Although the flat band feature and its corresponding high density of states can be observed using angle-resolved photoemission spectroscopy \cite{Lisi21,Stauber13,Jiang23,Li24} and scanning tunneling microscopy \cite{Tilak21}, respectively, systematic quantification of the degree of band flatness is still lacking. In this work, we consider different degrees of band flatness, ranging from narrow bands and nearly flat bands to ideal flat bands. TBG provides a highly tunable platform for engineering various degrees of band flatness.

In previous studies, most theoretical work on optical conductivity in twisted bilayer graphene (TBG) focused on gapless phases with larger twist angles \cite{Tabert13, Moon13a, Moon14, Stauber18} or near the twist angle in the presence of strain and small lattice relaxation \cite{Dai21, Wen21}. In this Letter, we propose using optical conductivity, along with its corresponding optical bound and the refined trace-determinant inequality (TDI) \cite{Chiu25}, to reveal the degree of band flatness and anisotropic quantum geometry \cite{supp00} by varying the twist angle and lattice relaxation. First, inspired by the experimental observation of the emergence of superconductivity in TBG, we show that the band flatness revealed by optical conductivity can be categorized into three regions and is consistent with experiments as the twist angle is varied \cite{Balents20}. Second, we calculate the optical bound, which is a new tool that can experimentally probe band flatness and quantum geometry, especially in the ideal flat band limit and in systems with broken time-reversal symmetry, by adjusting the lattice relaxation. Finally, we demonstrate the positive or negative definite properties of the Berry curvature in relation to the ideal flat band limit. Furthermore, we show the relationship between the vanishing of flat band velocities, emergent chiral symmetry \cite{Tarnopolsky19,Wang21,Song21}, and the saturation of the trace condition (TC) and the determinant condition (DC).

\textit{Continuum model of TBG-hBN heterostructures}.---The low-energy physics of TBG at small twist angles can be described by the Bistritzer-MacDonald (BM) model \cite{MacDonald11,Santos07}. In the layer basis, the effective Hamiltonian for valley $\xi=\pm$ can be written as
\begin{align}
	H_{\xi}(\mathbf{q}) &= \left(\begin{array}{cc}H_{b,\xi}(\mathbf{q}) & T_{\xi}(\mathbf{r}) \\ T^{\dagger}_{\xi}(\mathbf{r}) & H_{t,\xi}(\mathbf{q})\end{array}\right),
\end{align}
where $H_{b/t,\xi}(\mathbf{q})=\hbar\upsilon_{F}R(\mp\theta/2)\mathbf{q}\cdot(\xi\sigma_{x},\sigma_{y})$ is the Hamiltonian of bottom and top layer, which rotate about the $z$ axis mutually with a twist angle of $\pm\theta$. Here $\upsilon_{F}$ is the Fermi velocity of graphene, $R(\pm\theta/2)$ is the rotation matrix, and $\sigma_{x,y}$ is the Pauli matrix acting on the sublattice space of graphene. The wave vector $\mathbf{q} = \mathbf{k} - \mathbf{K}_{l,\xi}$ is defined with respect to the Dirac point $\mathbf{K}_{l,\xi}$ of graphene's Brillouin zone. The interlayer hopping is given by
\begin{align}\label{key}
	T_{\xi}(\mathbf{r})&=w_{1}	
	\begin{pmatrix}
		\kappa & 1 \\
		1 & \kappa  \\
	\end{pmatrix}e^{-i\xi\mathbf{q}_b\cdot\mathbf{r}}+
	\begin{pmatrix}
		\kappa & e^{-i\xi\frac{2\pi}{3}}  \\
		e^{i\xi\frac{2\pi}{3}} & \kappa  \\
	\end{pmatrix}e^{-i\xi\mathbf{q}_{tr}\cdot\mathbf{r}}   \\
	&+
	\begin{pmatrix}
		\kappa & e^{i\xi\frac{2\pi}{3}}  \\
		e^{-i\xi\frac{2\pi}{3}} & \kappa  \\
	\end{pmatrix}e^{-i\xi\mathbf{q}_{tl}\cdot\mathbf{r}},
\end{align}
where $w_0$ and $w_1$ are the AA and AB interlayer hopping parameters, respectively. We set $\kappa=w_0/w_1$ as the lattice relaxation parameter. The three momentum transfer vectors are given by $\mathbf{q}_b=k_{\theta}(0,-1)$, $\mathbf{q}_{tr}=k_{\theta}(\frac{1}{2},\frac{\sqrt{3}}{2})$, and $\mathbf{q}_{tl}=k_{\theta}(-\frac{1}{2},\frac{\sqrt{3}}{2})$, where $k_{\theta}=\frac{8\pi}{3\sqrt{3}d}\sin\frac{\theta}{2}$. Here $d=1.42\mathring{\text{A}}$ is the carbon-carbon bond length of graphene. We take $\hbar\upsilon_{F}=5944\ \text{eV}\mathring{\text{A}}$ and $w_1=110\ \text{eV}$ \cite{Song22} for all calculations. 

In order to study the quantum geometry of TBG, we need to break the gapless points, which can be achieved by introducing the hBN substrate \cite{Liu21,Gao24}. In the presence of hBN, the $C_{2z}$ symmetry is broken, and the top and bottom layer Hamiltonian becomes $H_{b/t,\xi}(\mathbf{q})+\Delta_{b/t}\sigma_z$, where $\Delta_{b/t}$ is the sublattice staggered potential. The role of the staggered potential is to open a gap to prevent the divergence of optical conductivity near zero frequency and to ensure that the quantum geometric tensor is well-defined.



\begin{figure*}[]
	\includegraphics[width=0.9\textwidth]{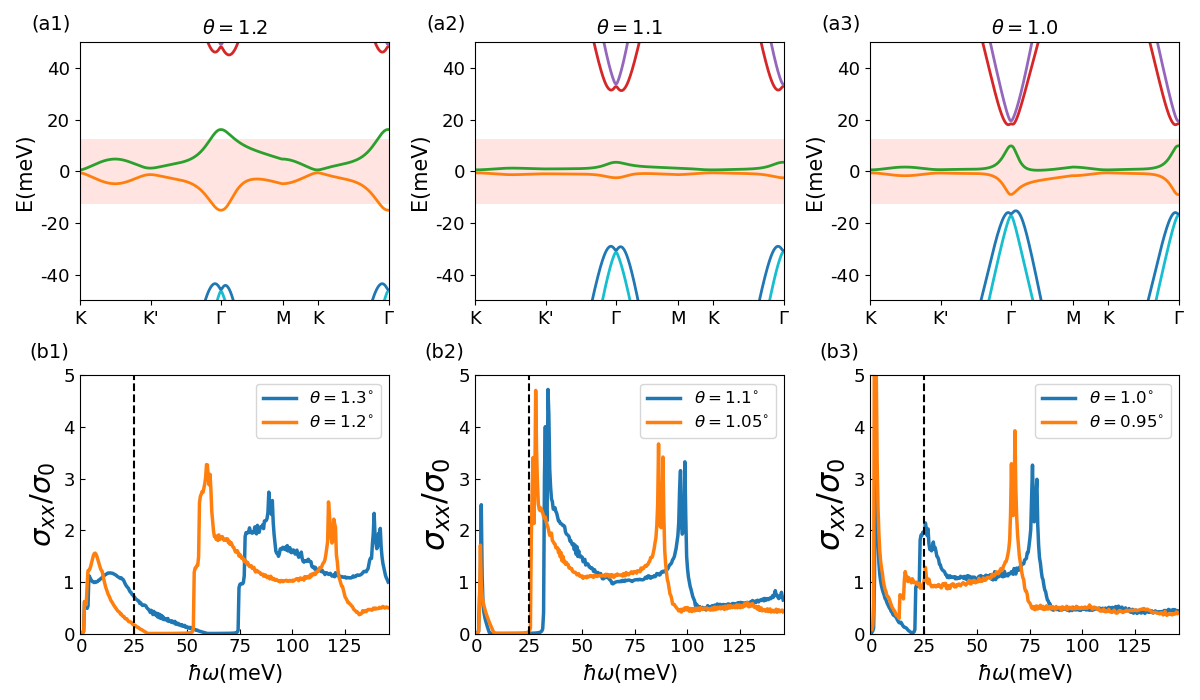}
	\caption{Band structure and optical conductivity as a function of the selected twist angle $\theta$. (a1-a3) Band structure of TBG with twist angle $\theta=1.2^{\circ},1.1^{\circ},1.0^{\circ}$. The color bands indicate the strength of Coulomb interactions extracted from STM experiments \cite{Xie19,Torma22}. (b1-b3) Optical conductivity with three regions of twist angles ($\theta_{c1}>\theta\ge1.2^{\circ}, 1.2^{\circ}>\theta>1.0^{\circ}, \theta_{c2}<\theta\le1.0^{\circ}$) in units of $\sigma_{0}=\frac{e^2}{h}$. Here we plot the optical conductivity with two twist angles in each region to illustrate the trend. The dashed black line marked in (b1-b3) represents the strength of Coulomb interaction extracted from STM data \cite{Xie19, Torma22}. Note that experimentally, superconductivity was found with twist angle in the region $1.2^{\circ}>\theta>1.0^{\circ}$ \cite{Balents20}, where the bandwidth, shown in the first zone bounded by zeros in the optical conductivity, is smaller than the strength Coulomb interaction. Here we set $\kappa = 0.8$, and $(\Delta_b, \Delta_t) = (5, 0)$ meV, which slightly opens a gap; the other parameters are the same as in the main text.}
\end{figure*}

\begin{figure*}[]
	\includegraphics[width=0.99\textwidth]{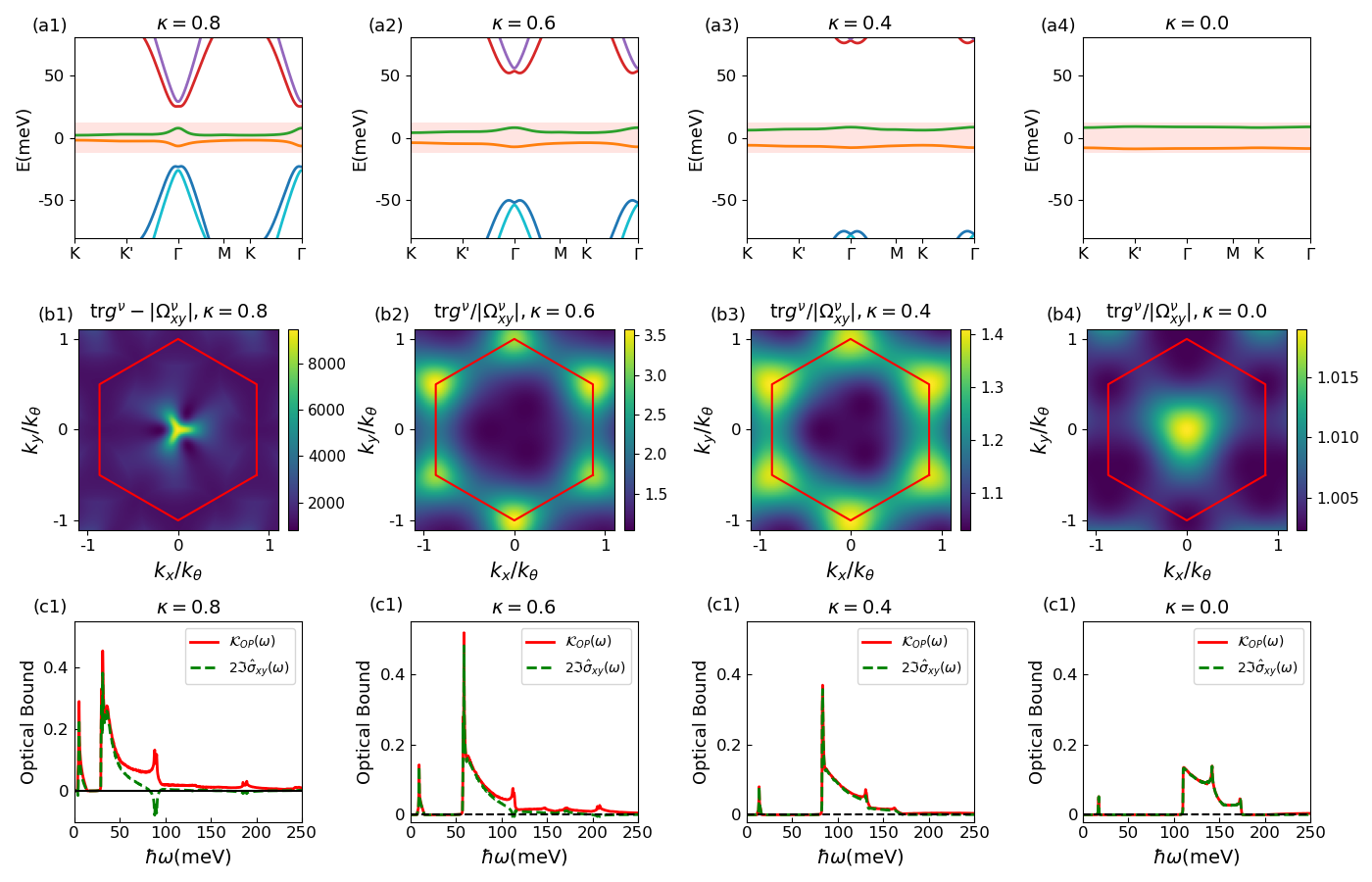}
	\caption{Band structure, trace condition, and optical bound (inequality) as a function of $\kappa=w_{0}/w_{1}$ with $\theta=1.05^{\circ}$. (a1-a4) Band structure of TBG with $\kappa=0.8,0.6,0.4,0.0$. The color bands indicate the strength of Coulomb interactions extracted from STM experiments \cite{Xie19,Torma22}. (b1-b4) Trace condition of the uppermost valence flat band with the corresponding values of $\kappa$. (c1-c4) Optical bound and imaginary part of generalized optical Hall conductivity with the corresponding value of $\kappa$, in units of $\sigma_{0}=\frac{e^2}{h}$. The red (green dashed) line represents the optical bound (imaginary part of generalized optical Hall conductivity). Note that because some values of $|\Omega_{xy}^{\nu}|$ are close to zero, we consider the subtraction form of the trace condition when $\kappa=0.8$.
	}
\end{figure*}

\textit{Optical Conductivity and Optical Bound}.---In this work, we focus on the linearly polarized optical conductivity, which can be expressed by the Kubo-Greenwood formula \cite{Kubo57,Greenwood58,Huhtinen23}:
\begin{align}
	\sigma_{ab}(\omega)=\frac{e^{2}}{i\hbar}\sum_{m\ne n}\int\frac{d^2\bf{k}}{(2\pi)^2}\frac{f_{mn}(\mathbf{k})}{E_{mn}(\mathbf{k})}  
	\frac{\mathcal{M}^{a}_{mn}(\mathbf{k})\mathcal{M}^{b}_{nm}(\mathbf{k})}{\hbar\omega+E_{mn}(\mathbf{k})+i\eta},
\end{align}
where the velocity matrix element $\mathcal{M}^{a}_{mn}(\mathbf{k})=\langle u_{m}(\mathbf{k})|\hbar\upsilon_{a}|u_{n}(\mathbf{k})\rangle$ and $a=x,y$.  $f_{n}(\mathbf{k})=1/[1+e^{(E_{n}(\mathbf{k})-\mu)/k_{B}T}]$ and $\upsilon_{a}=\frac{1}{\hbar}\frac{\partial H}{\partial k_{a}}$ are the Fermi-Dirac distribution function and velocity operator, respectively. We denote $f_{mn}(\mathbf{k})=f_{m}(\mathbf{k})-f_{n}(\mathbf{k})$ and $E_{mn}(\mathbf{k})=E_{m}(\mathbf{k})-E_{n}(\mathbf{k})$, where $H(\mathbf{k})|u_{n}(\mathbf{k})\rangle=E_{n}(\mathbf{k})|u_{n}(\mathbf{k})\rangle$. Here $\mu$ is the chemical potential, which we set to be located in the middle between the lowermost conduction flat band (denoted by $c$) and the uppermost valence flat band (denoted by $\nu$). In our calculation, we set a small phenomenological damping rate, $\eta = 0.1\ \text{meV}$, to reveal the features of the band structure and consider only a single valley contribution..


The low-energy region of the real part of the longitudinal optical conductivity provides information about the flat band, such as the bandwidth between two flat bands and the multi-gap features, which are the main interests of this work. To connect optical conductivity to quantum geometry, we assume there is a full gap between the occupied band and the unoccupied band. Therefore, the interband contribution to the Hermitian part of the optical conductivity in the clean limit is given by the following expression \cite{Souza08}:
\begin{align}
	\sigma_{\text{H},ab}(\omega)&=-\frac{\pi e^{2}}{\hbar}\sum_{m\in occ}\sum_{n\in unocc}\int\frac{d^2\bf{k}}{(2\pi)^2}\frac{f_{mn}(\mathbf{k})}{E_{mn}(\mathbf{k})}  \nonumber \\
	&\times\mathcal{M}^{a}_{mn}(\mathbf{k})\mathcal{M}^{b}_{nm}(\mathbf{k})\delta(\hbar\omega+E_{mn}(\mathbf{k})).
\end{align}

Using the identity between the velocity matrix element and the interband Berry connection, $\langle u_{m}(\mathbf{k})|\hbar\upsilon_{a}|u_n(\bm{k})\rangle=i E_{mn}r^{a}_{mn}(\bm{k})=iE_{mn}\langle u_{m}(\mathbf{k})|i\partial_{\bm{k}_a}|u_{n}(\mathbf{k})\rangle$ with $m\ne n$, the expression can be further rewritten in terms of the components of quantum geometric tensor, $Q^{mn}_{ab}(\bm{k})=r^{a}_{mn}(\bm{k})r^{b}_{nm}(\bm{k})=g^{mn}_{ab}(\bm{k})-\frac{i}{2}\Omega^{mn}_{ab}(\bm{k})$ \cite{Ahn20,Ahn22,supp0}, i.e.,
\begin{align}
	\sigma_{\text{H},ab}(\omega)&=\pi e^{2}\omega\int\frac{d^2\bf{k}}{(2\pi)^2}\sum_{m\in occ}\sum_{n\in unocc}Q^{mn}_{ab}(\bm{k})   \nonumber\\
	&\times\delta(\hbar\omega+E_{mn}(\mathbf{k})),
\end{align} 
where $g^{mn}_{ab}(\bm{k})$ and $\Omega^{mn}_{ab}(\bm{k})$ are the components of quantum metric and Berry curvature, respectively. $\sigma_{\text{H},ab}(\omega)=\Re\sigma_{aa}+i\Im\sigma_{ab}$ is the absorptive part of the optical conductivity \cite{Souza08}. Here we assume the system is at zero temperature.

Furthermore, to identify the degree of band flatness more precisely in optical response measurements, the following optical bound (inequality) is a useful tool \cite{Chiu25,supp1}: 
\begin{align}
	\mathcal{K}_{OP}(\omega)=\frac{2[\Re\sigma_{xx}(\omega)+\Re\sigma_{yy}(\omega)]}{\hbar\omega}\ge2\Im\hat{\sigma}_{xy}(\omega),
\end{align}
where
\begin{align}
	\Im\hat{\sigma}_{xy}(\omega)&=\frac{\pi e^{2}}{\hbar}\sum_{m\in occ}\sum_{n\in unocc}\int\frac{d^2\bf{k}}{(2\pi)^2}\Omega^{mn}_{xy}(\bm{k})  \nonumber\\
	&\times\delta(\hbar\omega+E_{mn}(\mathbf{k})),
\end{align}
which is the imaginary part of the generalized optical Hall conductivity \cite{Ebert96,Oppeneer98,Gradhand13,Onishi24} and can be obtained through the formula for the original optical Hall conductivity, i.e., $\Im\hat{\sigma}_{xy}(\omega)=\frac{-2\Im\sigma_{xy}(\omega)}{\hbar\omega}$. The circularly polarized optical conductivity, which has a close relationship with the vortexability condition \cite{Ledwith23}, can reveal similar physical properties (see the Supplemental Material (SM) for details \cite{supp}).

\textit{Results and Discussion}.---In the last section, we introduced the quantum geometry interpretation of optical conductivity and its associated optical bound, which is used for measuring the degree of band flatness and the anisotropic properties of a system. Now we demonstrate their application in TBG. Motivated by the experiments on superconductivity in TBG \cite{Balents20}, we divide the twist angle into three regions: $\theta_{c1}>\theta\ge1.2^{\circ}$, $1.2^{\circ}>\theta>1.0^{\circ}$ and $\theta_{c2}<\theta\le1.0^{\circ}$, and calculate the band structure and optical conductivity (see Fig. 1). The critical angles $\theta_{c1}$ and $\theta_{c2}$ can be determined by maintaining the corresponding features. The three regions exhibit entirely different tendencies. Importantly, the first nonzero zone, bounded by the zeros on the left-hand and right-hand sides of $\Re\sigma_{xx}(\omega)$, arises from the contributions of interband transitions between the nearly flat bands and provides the bandwidth of these bands. It is known that the bandwidth, which is (much) smaller than the electron interactions, is a crucial condition for the emergence of superconductivity \cite{Torma22}. Scanning tunneling microscopy experiments indicate that the Coulomb repulsion strength in TBG is around 25 meV \cite{Xie19,Torma22}, which we mark in Fig. 1(b1-b3) with a dash black line as a reference energy scale. With this interaction strength, in the region $1.2^{\circ}>\theta>1.0^{\circ}$ (see Fig. 1(a2,b2)), the bandwidth is smaller than 25 meV, making the emergence of superconductivity possible \cite{Cao18}. On the other hand, the isolated peak contributed by the optical transition between nearly flat bands is relatively narrow (see Fig. 1(b2)) compared to the twist angles outside this region (see Fig. 1(b1, b3)). This indicates that the effective mass of nearly flat band electrons is larger (more localized), according to the partial f-sum rule \cite{Hazra19,Verma21}.

In contrast, for the region $\theta_{c1}>\theta\ge1.2^{\circ}$ (see Fig. 1(a1,b1)), the bandwidth is larger than 25 meV. Compared to the crucial condition, the emergence of superconductivity is unlikely, which is consistent with the experimental results \cite{Balents20}. Furthermore, the isolated peak and the zone size between narrow band transitions are relatively low and much larger, respectively (see Fig. 1(b1)). This indicates that the corresponding bandwidth is larger and the effective mass of narrow band electrons is smaller (i.e., less localized).

In the region $\theta_{c2}<\theta\le1.0^{\circ}$ (see Fig. 1(a3,b3)), there is a different tendency: the gap between the nearly flat band and the dispersive band is very small, as revealed by the optical conductivity. The corresponding low-energy peak is much higher, and the zone size between the nearly flat band transitions overlaps with the transitions between the nearly flat band and the dispersive bands (see Fig. 1(b3)). This indicates that the nearly flat band has a smaller effective mass (i.e., less localized), and the corresponding gap between the nearly flat band and the dispersive band is small. As a result, flat band superconductivity is less likely since the electrons from the dispersive bands, which participate in or mediate the formation of Cooper pairs, have higher kinetic energy. However, there was an exceptional case found in two experimental groups \cite{Codecido19,Gao24}, which reported superconducting states in TBG with twist angles $\theta\approx0.93^{\circ}$ and $\theta=1.4^{\circ}$, respectively. This can be understood by the fact that the interaction strength can be adjusted by the experimental setup, such as dielectric media and substrates \cite{Liu21,Gao24}. As the superconducting gap is small ($<3\ \text{meV}$) in TBG \cite{Oh21,Koblischka24}, we expect that a similar feature will appear in the fully gapped superconducting state in the region $1.2^{\circ}>\theta>1.0^{\circ}$. 


\begin{figure}[]
	\includegraphics[width=0.49\textwidth]{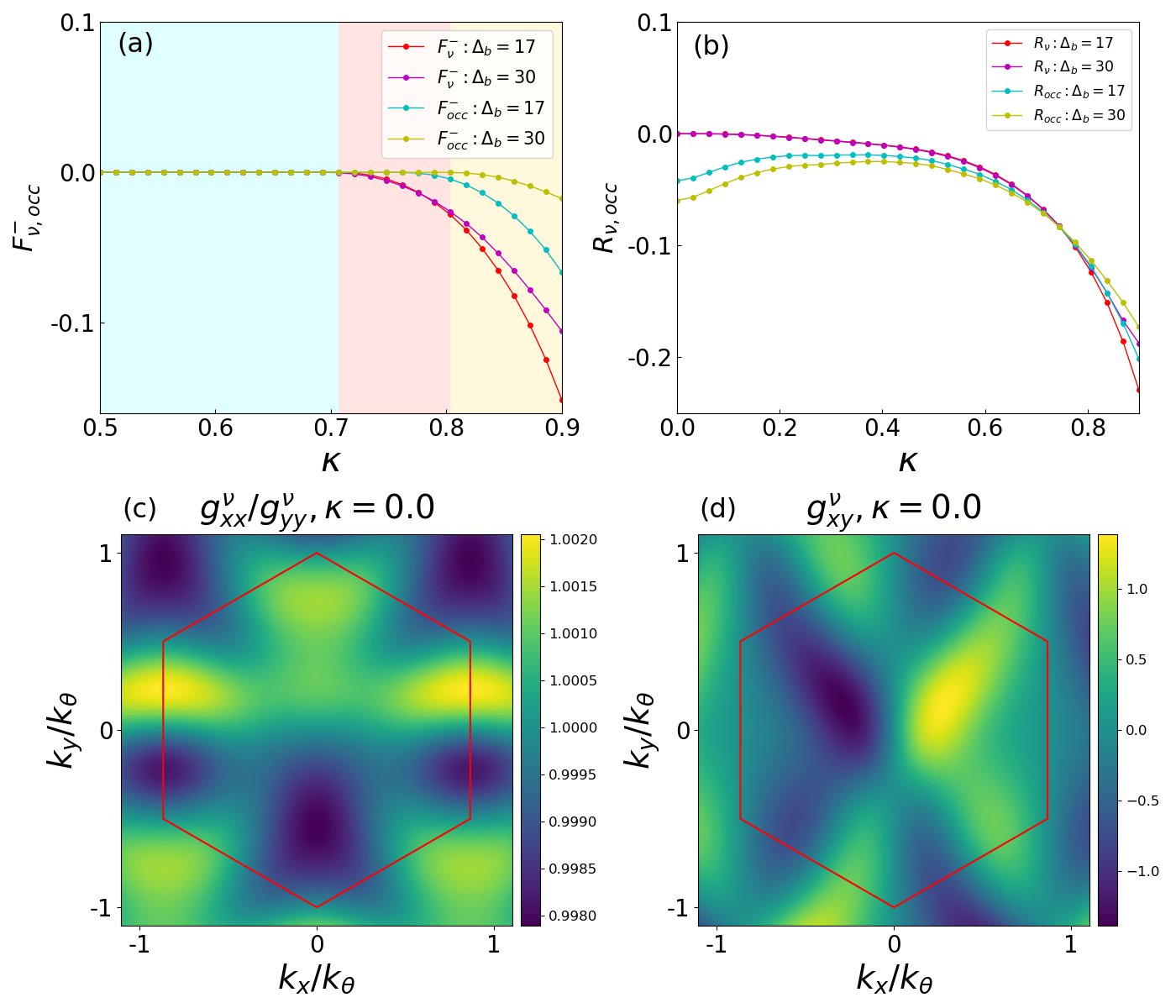}
	\caption{(a) The negative part of the Berry curvature of the valence flat and occupied band as a function of $\kappa$. Depending on whether $F^{-}_{\nu}$ and/or $F^{-}_{occ}$ reach zero, we divide the range of $\kappa$ into three colored regions. (b) The ratio of total amounts of the negative and positive components of Berry curvature of valence flat and occupied band as a function of $\kappa$. (c) The ratio of $g_{xx}^{\nu}$ and $g_{yy}^{\nu}$ with $\kappa=0$, is shown over the MBZ. (d) The off-diagonal component of the quantum metric, $g_{xy}^{\nu}$, with $\kappa=0$, is shown over the MBZ. Note that the values of $g_{xx}^{\nu}/g_{yy}^{\nu}$ and $g_{xy}^{\nu}$, close to one and zero respectively, imply the isotropic FCI phase. Here we set $\theta=1.05^{\circ}$. The evolution of the quantum metric as a function of $\kappa$ is shown in the SM \cite{supp}.
	}
\end{figure}

To realize the FCI phases, nearly flat bands are not sufficient; even flatter bands are needed, ideally approaching an ideal flat band \cite{Ledwith20, Wang21, Wang23, Estienne23, Ledwith23, Liu25}. By tuning the lattice relaxation, we demonstrate the evolutionary process of band flatness from a nearly flat band to an ideal flat band through the optical bound inequality \cite{Chiu25}. In Fig. 2(c1-c4), we show the optical bound for four selected lattice relaxations, along with the TC of the uppermost valence band (see Fig. 2(b1-b4)) and the corresponding band structure (see Fig. 2(a1-a4)). As shown in Fig. 2, there is a mutual tendency: as the value of $\kappa$ decreases, the first peak of the optical bound consistently becomes lower, the TC tends to saturate, and the band gap increases. Furthermore, the values of $\mathcal{K}_{OP}(\omega)$ and $2\Im\hat{\sigma}_{xy}(\omega)$ become increasingly close as $\kappa$ decreases, overlapping in the isotropic ideal flat band limit. The complete overlap between the optical bound and the generalized optical Hall conductivity is a signal of the saturation of TC. As a first step toward saturation, the negative value of $2\Im\hat{\sigma}_{xy}(\omega)$ vanishes and becomes entirely positive (or negative) near a critical value, specifically $\kappa\approx0.6$ in our case. This is constrained by the refined TDI \cite{Chiu25}:
\begin{align}
	\label{eq:e1}
	\mathrm{tr} g({\bf k}) \ge 2\sqrt{\det(g({\bf k}))} \ge \overline{\Omega}_{ab}({\bf k})\ge|\Omega_{ab}({\bf k})|,
\end{align}
where $\overline{\Omega}_{ab}({\bf k})$ is the maximal Berry curvature \cite{Chiu25} that characterizes the finer structure of the topology of a system; $\overline{\Omega}_{ab}({\bf k})$ and $\Omega_{ab}({\bf k})$ become the same in the ideal flat band limit.

As constrained by the refined TDI, the band-revolved components of Berry curvature are (semi-)positive or (semi-)negative definite in the ideal flat band limit. This property is treated as an assumption in the definition of an ideal flat band \cite{Wang21}. Recall that the expression of occupied band Berry curvature is $\Omega_{xy}(\bm{k})=-2\sum_{m\in occ}\sum_{n\in unocc}\Im(r^{x}_{mn}(\bm{k})r^{y}_{nm}(\bm{k}))$. We define the amount of negative part of Berry curvature as 
$F^{-}_{occ}=\int \frac{d\mathbf{k}}{(2\pi)^2}\min(\Omega_{xy}(\bm{k}),0)$. For comparing the value of the total amount of negative and positive component of Berry curvature, we use their ratio, $R_{occ}=\widehat{F}^{-}_{occ}/\widehat{F}^{+}_{occ}$, where
$\widehat{F}^{-}_{occ}=\int \frac{d\mathbf{k}}{(2\pi)^2}\sum_{m\in occ}\sum_{n\in unocc}\min(\Omega^{mn}_{xy}(\bm{k}),0)$ and $\widehat{F}^{+}_{occ}=\int \frac{d\mathbf{k}}{(2\pi)^2}\sum_{m\in occ}\sum_{n\in unocc}\max(\Omega^{mn}_{xy}(\bm{k}),0)$ are the total amount of negative and positive component of Berry curvature, respectively. Here $\Omega^{mn}_{xy}(\bm{k})=-2\Im(r^{x}_{mn}(\bm{k})r^{y}_{nm}(\bm{k}))$. In the case of valence flat band Berry curvature, one can replace $\sum_{m\in occ}\sum_{n\in unocc}$ with $\sum_{n\ne \nu}$ to obtain the corresponding formulas for $F^{-}_{\nu}$ and $R_{\nu}$. In Fig. 3(a), it is shown that both $F^{-}_{\nu}$ and $F^{-}_{occ}$ saturate to zero after $\kappa=0.7$, which is consistent with the calculations in \cite{Xie21,Parker21}, indicating the beginning of the FCI phase. For the occupied band Berry curvature, the saturation occurs slightly faster. Additionally, we can see that a larger $\Delta$ only makes a difference before saturation and is significant when there is no lattice relaxation. The ratios $R_{\nu}$ and $R_{occ}$ further reveal the (semi-)positive and (semi-)negative definite property of Berry curvature in the ideal flat band limit. As shown in Fig. 3(b), $R_{\nu}$ saturates to zero. It indicates the positive and negative definite property of valence flat band of Berry curvature.
In contrast, $R_{occ}$ is slightly away from zero because not all occupied bands are flat. In Fig. 3(c,d), it is shown that the ratio $g_{xx}^{\nu}/g_{yy}^{\nu}$ approaches one and $g_{xy}^{\nu}$ approaches zero, respectively, in the chiral limit. This is closely related to the saturation of TC, which is studied below.

Now we demonstrate the relationship between the vanishing of flat band velocities, emergent chiral symmetry, and the saturation of TC and DC. The (semi-)positive or (semi-)negative definite property of Berry curvature suggests that it may be related to certain conditions and symmetries. Firstly, we observe that the first inequality, $\mathrm{tr} g({\bf k}) \ge 2\sqrt{\det(g({\bf k}))}$ shows that it saturates only when $g_{xx}({\bf k})=g_{yy}({\bf k})$ and $g_{xy}({\bf k})=0$. Using this constraint for the second inequality, $2\sqrt{\det(g({\bf k}))} \ge|\Omega_{ab}({\bf k})|$, the saturation of TC reduces to $g_{xx}({\bf k})\ge|\Omega_{xy}({\bf k})|/2$, which is a special case of the ideal flat band condition proposed in Refs. \cite{Wang21,Wang23}. Next, we are going to show that saturation is enforced by both the vanishing of flat band velocities at the magic angle ($\theta=\theta_c$) and emergent chiral symmetry ($\kappa=0$) \cite{Tarnopolsky19,Wang21,Song21}. In the presence of emergent chiral symmetry, the velocity operators $\upsilon_x$ and $\upsilon_y$ are related \cite{Sheffer23}, specifically $\upsilon_y(\mathbf{k},\theta)=-iS\upsilon_x(\mathbf{k},\theta)$, where $S$ is the operator of emergent chiral symmetry \cite{Song21,Sheffer23}. Here we emphasize twist angle $\theta$ as a tuning parameter. When projecting onto the two flat band subspace \cite{Sheffer23}, we obtain,
\begin{align}
	|\upsilon^x_{\nu c}(\mathbf{k},\theta_c)|^2+|\upsilon^y_{\nu c}(\mathbf{k},\theta_c)|^2=|-2\Im(\upsilon^x_{\nu c}(\mathbf{k},\theta_c)\upsilon^y_{c\nu}(\mathbf{k},\theta_c))|,
\end{align}	
where $\upsilon^a_{\nu c}(\mathbf{k},\theta_c)$ is the flat band component of the velocity operator. Using this relation, the saturation of single-band and multi-band TC can be achieved under certain conditions(see SM \cite{supp} for details). 


In anisotropic systems, TC is no longer saturated; instead, DC can be saturated. This is ensured by the condition for saturation of the Cauchy-Schwarz inequality for each occupied band's complex vector \cite{Ozawa21}: $(1-P(\mathbf{k}))|\partial_{k_x} u_m(\mathbf{k})\rangle=c(1-P(\mathbf{k}))|\partial_{k_y}u_m(\mathbf{k})\rangle$, which provides the relations between the elements of the quantum geometric tensor (see SM \cite{supp}). Here $c=c_1+ic_2$ is a complex number that takes values in the upper (lower) half of the complex plane, $\mathbb{H}=\{c\in \mathbb{C}:\Im(c)>0\}$ or $\{c\in \mathbb{C}:\Im(c)<0\})$, if the Berry curvature is (semi-)positive ((semi-)negative) definite \cite{supp,supp2}. As a result, quantum metric and Berry curvature are related by a "saturation matrix", i.e.,
\begin{align}
	\begin{pmatrix}
		g_{xx}(\mathbf{k}) & g_{xy}(\mathbf{k}) \\
		g_{xy}(\mathbf{k}) & g_{yy}(\mathbf{k}) \\
	\end{pmatrix}=
     \begin{pmatrix}
     	0 & \frac{\Omega_{xy}(\mathbf{k})}{2} \\
     	-\frac{\Omega_{xy}(\mathbf{k})}{2} & 0 \\
     \end{pmatrix}
     \begin{pmatrix}
     	-\frac{c_1}{c_2} & -\frac{1}{c_2} \\
     	\frac{c_1^2+c_2^2}{c_2} & \frac{c_1}{c_2} \\
     \end{pmatrix}.      	    
\end{align}
Note that the "saturation matrix" can be interpreted as the almost complex structure when using the description of Kähler geometry \cite{Mera21a,Mera21b,Mera24,supp3}, and is related to lattice deformation \cite{Avron95,Levay95,Levay97,Tokatly07}. When setting $c_1=0$ and $c_2=1$, one can reduce it to the isotropic case. Remarkably, the saturation constant $c$ can be measured using the Souza-Wilkens-Martin sum rule \cite{Souza00} or the generalized optical weight \cite{Onishi24}: $c_1=\frac{\sqrt{4W_{xx}W_{yy}-\mathcal{C}}}{2W_{yy}}$ and 
$c_2=\frac{\mathcal{C}}{2W_{yy}}$, where $W_{aa}=\frac{\hbar}{e^2}\int^{\infty}_{0}\frac{2\Re\sigma_{aa}(\omega)}{\omega}d\omega$ and $\mathcal{C}=-\frac{\hbar}{e^2}\int^{\infty}_{0}\frac{4\Im\hat{\sigma}_{xy}(\omega)}{\omega}d\omega$ (see SM \cite{supp}). In other words, the saturation constant $c$, which represents the intrinsic anisotropy of a system \cite{Haldane11,Haldane11b,Qiu12,BYang12,BYang17} or its lattice deformation by applying strain \cite{Read09,Read11,Bagrov17,Paiva25}, can be measured experimentally when the system is in a FCI phase or hosts ideal flat bands. On the other hand, in anisotropic systems, we always have $g_{xx}(\mathbf{k})+g_{yy}(\mathbf{k})>|\Omega_{xy}(\mathbf{k})|$. As a result, the non-vanishing difference between the optical bound and the imaginary part of the generalized optical Hall conductivity also provides an anisotropic signature in the FCI phases.

\textit{Conclusions}.---We have demonstrated how both the optical conductivity and the optical bound can reveal the degree of band flatness and the anisotropic quantum geometry in TBG. Moreover, we have shown the relationship between the vanishing of flat band velocities, emergent chiral symmetry, and the saturation of the trace and determinant conditions for the realization of the FCI phases, as well as their possible anisotropic signatures in optical conductivity measurements. Our work can be directly applied to other moiré superlattice materials and kagome materials, providing a unified perspective for studying flat band systems.


\textit{Acknowledgments}
The author thanks Bruno Mera for pointing out the relationship between the 'saturation matrix' and the (almost) complex structure after completing the first version of the manuscript. The author also thanks Yarden Sheffer and Raquel Queiroz for explaining their work. The author was supported by the postdoctoral fellowship of the Ministry of Science and Technology (MOST) in Taiwan under grant no. MOST 111-2636-M-007-003 and 111-2811-M-004-001.

\appendix

\bibliography{ref_CD}

\pagebreak
\newpage

\thispagestyle{empty}
\mbox{}
\pagebreak
\newpage
\onecolumngrid
\begin{center}
	\textbf{\large Supplemental Materials: Optical Signatures of Band Flatness and Anisotropic Quantum Geometry in Magic-Angle Twisted Bilayer Graphene}
\end{center}

\author{Pok Man Chiu}\email{pokman2011@gmail.com}
\affiliation{Graduate Institute of Applied Physics, National Chengchi University, Taipei City 11605, Taiwan}
\affiliation{Department of Physics, National Tsing Hua University, Hsinchu 30013, Taiwan}

\setcounter{equation}{0}
\setcounter{figure}{0}
\setcounter{table}{0}
\setcounter{page}{1}
\makeatletter
\renewcommand{\theequation}{S\arabic{equation}}
\renewcommand{\thefigure}{S\arabic{figure}}
\renewcommand{\bibnumfmt}[1]{[S#1]}
\renewcommand{\citenumfont}[1]{S#1}

\onecolumngrid

In this supplementary material, we include (a) the evolution of the quantum metric with varying lattice relaxation, (b) detailed derivations of the relationship between the vanishing of flat band velocities, emergent chiral symmetry, and the saturation of the trace and determinant conditions, and (c) the relationship between the multi-band vortexability condition and the circular dichroism in TBG.

\subsection{1: Evolution of the Quantum Metric with Varying Lattice Relaxation}

To understand how the emergent chiral symmetry enforces the saturation of the trace condition, we first demonstrate the evolution of the quantum metric and trace condition for both the valence flat band and all occupied bands by varying lattice relaxation. As shown in Fig. S1, when we vary $\kappa$ from 0.8 to 0.0, the Abelian quantum metric ratio $g_{xx}^{\nu}/g_{yy}^{\nu}$ and $g_{xy}^{\nu}$ approach one and zero, respectively. This implies a relationship between emergent chiral symmetry and saturation of trace condition of the valence flat band. In Fig. S2, however, because not all occupied bands are ideal flat bands, the non-Abelian quantum metric ratio $g_{xx}/g_{yy}$, and $g_{xy}$ deviate slightly from one and noticeably from zero, respectively. On the other hand, since the Chern number of all occupied bands is one; the trace condition for all occupied bands cannot be saturated. It implies Chern number may be not a suitable topological invariant for all occupied bands of TBG.

\begin{figure*}[]
	\includegraphics[width=0.99\textwidth]{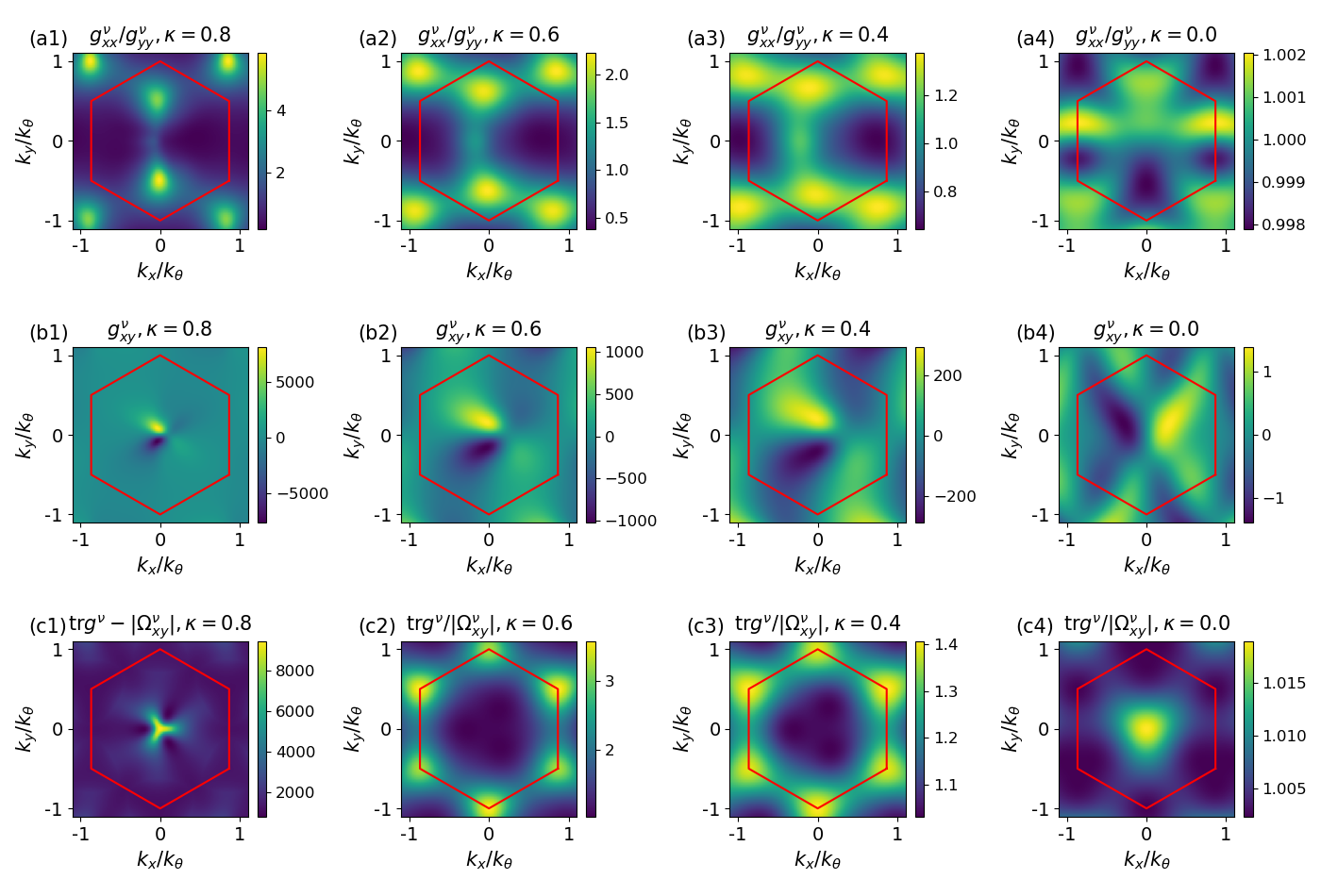}
	\caption{Evolution of the valence flat band's Abelian quantum metric and its trace condition as varying lattice relaxation, with $\theta=1.05^{\circ}$. (a1-a4) Evolution of the Abelian quantum metric ratio, $g_{xx}^{\nu}/g_{yy}^{\nu}$. (b1-b4) Evolution of the off-diagonal Abelian quantum metric element, $g_{xy}^{\nu}$. (c1-c4) Evolution of the trace condition of the valence flat band.  Note that because some values of $|\Omega_{xy}^{\nu}|$ are close to zero, we consider the subtraction form of the trace condition when $\kappa=0.8$.
	}
\end{figure*}

\begin{figure*}[]
	\includegraphics[width=0.99\textwidth]{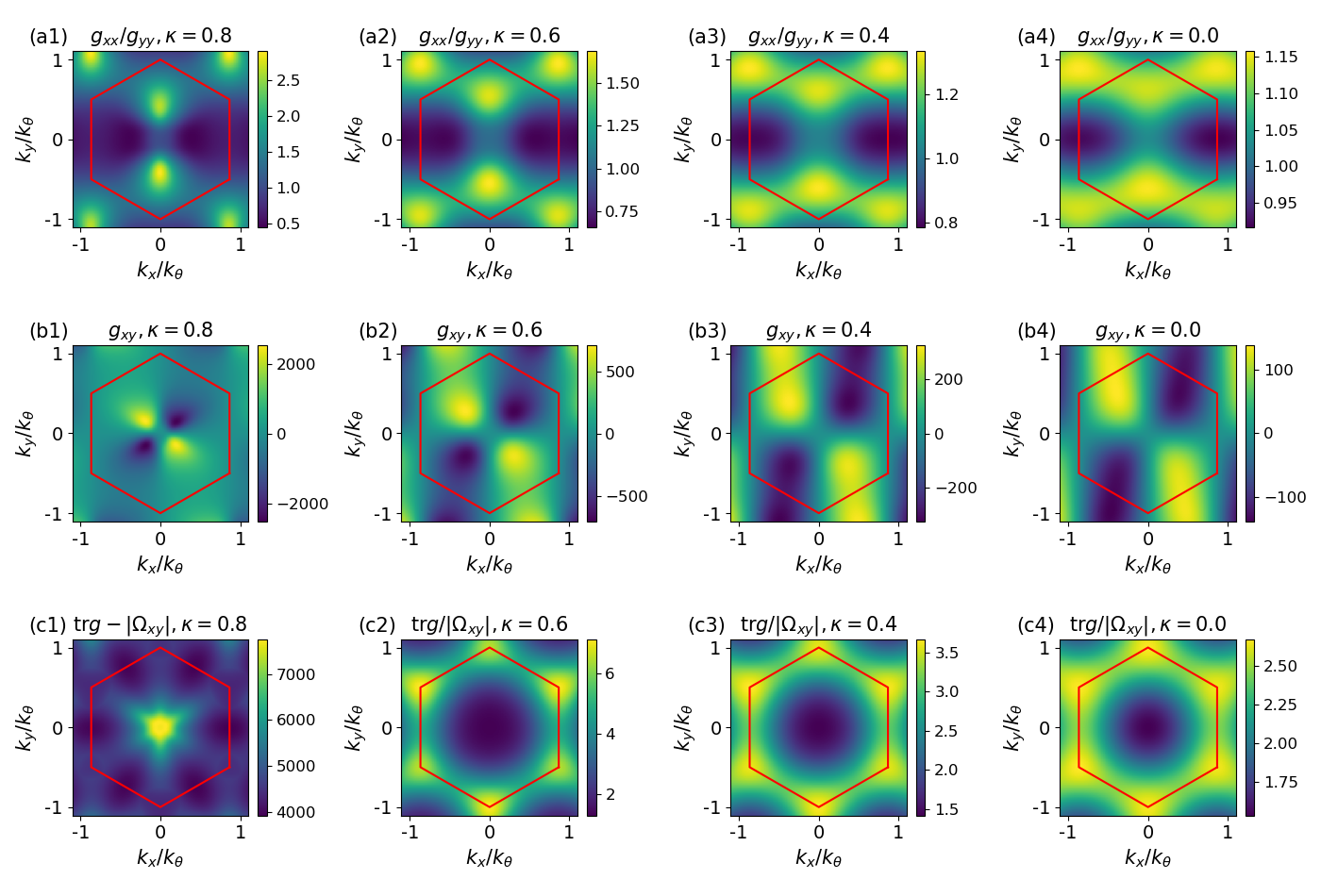}
	\caption{Evolution of the quantum metric of all occupied bands and its trace condition as varying lattice relaxation, with $\theta=1.05^{\circ}$. (a1-a4) Evolution of the non-Abelian quantum metric ratio, $g_{xx}/g_{yy}$. (b1-b4) Evolution of off-diagonal non-Abelian quantum metric element, $g_{xy}$. (c1-c4) Evolution of the trace condition of all occupied bands. Note that because some values of $|\Omega_{xy}^{\nu}|$ are close to zero, we consider the subtraction form of the trace condition when $\kappa=0.8$.
	}
\end{figure*}

\subsection{2: Relationship Between the Vanishing of Flat Band Velocities, Emergent Chiral Symmetry, and the Saturation of Trace and Determinant Conditions}

In this section, we provide detailed derivations of the relationship between the vanishing of flat band velocities, emergent chiral symmetry, and the saturation of the trace and determinant conditions. From the results in Ref. \cite{Sheffer23}, we know that the emergent chiral symmetry in TBG establishes a connection between the velocity operators $\upsilon_x$ and $\upsilon_y$, i.e.,
\begin{align}
	&\upsilon_y=-iS\upsilon_x,
\end{align}
where $S$ is the emergent chiral symmetry when $\kappa=0$, and $\upsilon_F$ is the Fermi velocity. We argue that, although the term for the sublattice staggered potential breaks the emergent chiral symmetry, the relationship between $\upsilon_x$ and $\upsilon_y$ is still preserved because it vanishes after differentiation. Crucially, the emergent chiral symmetry provides an approach establishing a relationship between the elements of the quantum geometric tensor. To start, we rewrite Eq. (1S) into two equivalent forms:
\begin{align}
	&\upsilon_x\upsilon_y=-i\upsilon_xS\upsilon_x,  \\
	&\upsilon_y\upsilon_x=i\upsilon_yS^{-1}\upsilon_y.
\end{align}
When projecting onto the subspace spanned by the two flat bands near zero energy \cite{Sheffer23}, Eqs. (S2) and (S3) becomes
\begin{align}
	&\rho(\upsilon_x)\rho(\upsilon_y)=-i\rho(\upsilon_x)\rho(S)\rho(\upsilon_x),  \\
	&\rho(\upsilon_y)\rho(\upsilon_x)=i\rho(\upsilon_y)\rho(S^{-1})\rho(\upsilon_y).
\end{align}
Since $S$ is a diagonal operator with $1$ and $-1$ as the nonzero entries \cite{Sheffer23}, however, we choose a more general form, i.e., $\begin{pmatrix} s_1 & 0 \\ 0  & s_2  \end{pmatrix}$, where the values of $s_1$ and $s_2$ will be determined later. Eqs. (S4) and (S5) become
\begin{align}
	&\begin{pmatrix}
	\upsilon^x_{\nu\nu}\upsilon^y_{\nu\nu}+\upsilon^x_{\nu c}\upsilon^y_{c\nu} & \upsilon^x_{\nu\nu}\upsilon^y_{\nu c}+\upsilon^x_{\nu c}\upsilon^y_{cc} \\
	\upsilon^x_{c\nu}\upsilon^y_{\nu \nu}+\upsilon^x_{cc}\upsilon^y_{c\nu} & \upsilon^x_{c\nu}\upsilon^y_{\nu c}+\upsilon^x_{c c}\upsilon^y_{cc} \\
    \end{pmatrix}  
    =-i
	\begin{pmatrix}
	s_1|\upsilon^x_{\nu\nu}|^2+s_2|\upsilon^x_{\nu c}|^2 & 	s_1|\upsilon^x_{\nu\nu}|^2+s_2\upsilon^x_{\nu c}\upsilon^x_{\nu\nu} \\
	s_1\upsilon^x_{c\nu}\upsilon^x_{cc}+s_2\upsilon^x_{\nu\nu}\upsilon^x_{\nu c} & +s_1|\upsilon^x_{c\nu}|^2+s_2|\upsilon^x_{cc}|^2 \\
    \end{pmatrix}
\end{align}
and 
\begin{align}
    &\begin{pmatrix}
    	\upsilon^y_{\nu\nu}\upsilon^x_{\nu\nu}+\upsilon^y_{\nu c}\upsilon^x_{c\nu} & \upsilon^y_{\nu\nu}\upsilon^x_{\nu c}+\upsilon^x_{\nu c}\upsilon^y_{cc} \\
    	\upsilon^y_{c\nu}\upsilon^x_{\nu \nu}+\upsilon^y_{cc}\upsilon^x_{c\nu} & \upsilon^y_{c\nu}\upsilon^x_{\nu c}+\upsilon^y_{c c}\upsilon^x_{cc} \\
    \end{pmatrix}  
    =i
    \begin{pmatrix}
    	\frac{1}{s_1}|\upsilon^y_{\nu\nu}|^2+\frac{1}{s_2}|\upsilon^y_{\nu c}|^2 & 	\frac{1}{s_1}|\upsilon^y_{\nu\nu}|^2+\frac{1}{s_2}\upsilon^y_{\nu c}\upsilon^y_{\nu\nu} \\
    	\frac{1}{s_1}\upsilon^y_{c\nu}\upsilon^y_{cc}+\frac{1}{s_2}\upsilon^y_{\nu\nu}\upsilon^y_{\nu c} & +\frac{1}{s_1}|\upsilon^y_{c\nu}|^2+\frac{1}{s_2}|\upsilon^y_{cc}|^2 \\
    \end{pmatrix},
\end{align}
respectively. In the following, we denote $\upsilon^{ab}_{mn}=\upsilon^{ab}_{mn}(\mathbf{k},\theta)$ to emphasize the twist angle as a tuning parameter. By comparing the left corner element of Eqs. (S6) and (S7) for both sides at the magic angle, i.e., $\theta=\theta_c$, we have
\begin{align}
    &i\upsilon^x_{\nu c}(\mathbf{k},\theta_c)\upsilon^y_{c\nu}(\mathbf{k},\theta_c)=s_2|\upsilon^x_{\nu c}(\mathbf{k},\theta_c)|^2
\end{align}
and
\begin{align}
	&-i\upsilon^y_{\nu c}(\mathbf{k},\theta_c)\upsilon^x_{c\nu}(\mathbf{k},\theta_c)=\frac{1}{s_2}|\upsilon^y_{\nu c}(\mathbf{k},\theta_c)|^2,
\end{align}
respectively. Here, we use the fact that $\upsilon^{a}_{\nu\nu}(\mathbf{k},\theta_c)=0=\upsilon^{a}_{cc}(\mathbf{k},\theta_c)$. Summing Eqs. (S8) and (S9), we finally obtain
\begin{align}
	&-2\Im(\upsilon^x_{\nu c}(\mathbf{k},\theta_c)\upsilon^y_{c\nu}(\mathbf{k},\theta_c))=s_2|\upsilon^x_{\nu c}(\mathbf{k},\theta_c)|^2+\frac{1}{s_2}|\upsilon^y_{\nu c}(\mathbf{k},\theta_c)|^2.
\end{align}
Here, $s_2$ is equal to $\pm1$ in the case of the emergent chiral symmetry, which makes $\upsilon^a_{\nu c}(\mathbf{k},\theta_c)\upsilon^b_{c\nu}(\mathbf{k},\theta_c)$ purely imaginary. Remarkably, the case of $s_2\ne\pm1$ implies a generalized TC \cite{ZLiu25,Paiva25}. 
	
Next, we derive the saturation of TC for the valence flat band. Since we can project the velocity operator onto the subspace of any pair of bands, we consider terms such as $|\upsilon^a_{\nu n}(\mathbf{k},\theta_c)|^2$ and $\Im(\upsilon^x_{\nu n}(\mathbf{k},\theta_c)\upsilon^y_{n\nu}(\mathbf{k},\theta_c))$, where $n\ne\nu$. By multiplying both sides by $\hbar^2/(E_{\nu}(\mathbf{k},\theta_c)-E_n(\mathbf{k},\theta_c))^2$ and taking the summation $\sum_{n\ne \nu}$ for Eq. (S10), we obtain the saturation of TC for the valence flat band, i.e.,
\begin{equation}
	0=\sum_{n\ne \nu}\frac{|\hbar\upsilon^x_{\nu n}(\mathbf{k},\theta_c)|^2+|\hbar\upsilon^y_{\nu n}(\mathbf{k},\theta_c)|^2-|-2\Im(\hbar\upsilon^x_{\nu n}(\mathbf{k},\theta_c)\hbar\upsilon^y_{n\nu }(\mathbf{k},\theta_c))|}{(E_{\nu}(\mathbf{k},\theta_c)-E_n(\mathbf{k},\theta_c))^2}=g^{\nu}_{xx}(\mathbf{k},\theta_c)+g^{\nu}_{yy}(\mathbf{k},\theta_c)-|\Omega^{\nu}_{xy}(\mathbf{k},\theta_c)|.
\end{equation}
Here, we set $s_2=1$. Using a similar method, the saturation of TC for the conduction flat band can also be obtained. Lastly, we consider the case of all occupied bands. Since we can project the velocity operator onto the subspace of any pair of occupied bands, which is not necessarily formed by flat bands, generally we have
\begin{align}
	s_1|\upsilon^x_{mm}(\mathbf{k},\theta_c)|^2+\frac{1}{s_1}|\upsilon^y_{mm}(\mathbf{k},\theta_c)|^2+s_2|\upsilon^x_{mn}(\mathbf{k},\theta_c)|^2+\frac{1}{s_2}|\upsilon^y_{mn}(\mathbf{k},\theta_c)|^2=-2\Im(\upsilon^x_{mn}(\mathbf{k},\theta_c)\upsilon^y_{nm}(\mathbf{k},\theta_c)).
\end{align}	 
Similarly, by multiplying both sides of Eq. (S13) by $\hbar^2/(E_m(\mathbf{k},\theta_c)-E_n(\mathbf{k},\theta_c))^2$ and taking the summation $\sum_{m\in occ}\sum_{n\in unocc}$, we can obtain
\begin{align}
	\sum_{m\in occ}\sum_{n\in unocc}\frac{|\hbar\upsilon^x_{mn}(\mathbf{k},\theta_c)|^2+|\hbar\upsilon^y_{mn}(\mathbf{k},\theta_c)|^2}{(E_m(\mathbf{k},\theta_c)-E_n(\mathbf{k},\theta_c))^2}\approx\sum_{m\in occ}\sum_{n\in unocc}\frac{|-2\Im(\hbar\upsilon^x_{mn}(\mathbf{k},\theta_c)\hbar\upsilon^y_{nm}(\mathbf{k},\theta_c))|}{(E_m(\mathbf{k},\theta_c)-E_n(\mathbf{k},\theta_c))^2}.
\end{align}
Here, we take $s_1=-1$ and $s_2=1$, and assume that $|\upsilon^x_{mm}(\mathbf{k},\theta_c)|^2$ and $|\upsilon^y_{mm}(\mathbf{k},\theta_c)|^2$ are small at the magic angle. In other words, the TC for all occupied bands cannot be saturated unless the velocities of all occupied bands, $|\upsilon^x_{mm}(\mathbf{k},\theta_c)|^2$ and $|\upsilon^y_{mm}(\mathbf{k},\theta_c)|^2$, are zero at the magic angle. That is, in general, we will have $g_{xx}(\mathbf{k},\theta_c)+g_{yy}(\mathbf{k},\theta_c)\ge|\Omega_{xy}(\mathbf{k},\theta_c)|$. However, when the velocities of all occupied (flat) bands are vanishing, the multi-band TC can be recovered. We have demonstrated how the vanishing of flat band velocities and the emergent chiral symmetry enforce the saturation of single-band and multi-band TC, which are only applicable to isotropic systems.

In anisotropic systems, TC is no longer saturated; instead, DC can be saturated \cite{Claassen15}. This is ensured by the saturation condition of the Cauchy-Schwarz inequality \cite{Ozawa21}: 
\begin{equation}
	(1-P(\mathbf{k},\theta_c))|\partial_{k_x} u_m(\mathbf{k},\theta_c)\rangle=c(1-P(\mathbf{k},\theta_c))|\partial_{k_y}u_m(\mathbf{k},\theta_c)\rangle,
\end{equation}
which provides the relationship between the elements of the quantum geometric tensor, where $P(\mathbf{k},\theta_c)=\sum_{n\in occ}|u_n(\mathbf{k},\theta_c)\rangle\langle u_{n}(\mathbf{k},\theta_c)|$ is the projection operator. Here $c$ is a complex number and can be considered a variational parameter for certain observables \cite{Claassen15}. By multiplying $\langle \partial_{k_a}u_{m}(\mathbf{k},\theta_m)|$ on both sides of Eq. (S18), and taking the summation over all unoccupied bands, we have
\begin{align}
	\sum_{m\in unocc}\langle \partial_{k_x}u_{m}(\mathbf{k},\theta_c)|(1-P(\mathbf{k},\theta_c))|\partial_{k_x} u_m(\mathbf{k},\theta_c)\rangle=c\sum_{m\in unocc}\langle \partial_{k_x}u_{m}(\mathbf{k,\theta_c})|(1-P(\mathbf{k},\theta_c))|\partial_{k_y}u_m(\mathbf{k},\theta_c)\rangle,  \\
	\sum_{m\in unocc}\langle \partial_{k_y}u_{m}(\mathbf{k},\theta_c)|(1-P(\mathbf{k},\theta_c))|\partial_{k_x} u_m(\mathbf{k},\theta_c)\rangle=c\sum_{m\in unocc}\langle \partial_{k_y}u_{m}(\mathbf{k},\theta_c)|(1-P(\mathbf{k},\theta_c))|\partial_{k_y}u_m(\mathbf{k},\theta_c)\rangle.	
\end{align}
Hence we obtain
\begin{align}
	g_{xx}(\mathbf{k},\theta_c)&=c(g_{xy}(\mathbf{k},\theta_c)-\frac{i}{2}\Omega_{xy}(\mathbf{k},\theta_c)),  \\
	g_{yy}(\mathbf{k},\theta_c)&=\frac{1}{c}(g_{xy}(\mathbf{k},\theta_c)+\frac{i}{2}\Omega_{xy}(\mathbf{k},\theta_c)).	
\end{align}
By directly comparing the real and imaginary parts of Eqs. (S21) and (S22), and setting $c=i$, it reduces to the saturation of TC in the isotropic case, i.e., 
\begin{align}
	g_{xx}(\mathbf{k},\theta_c)&=\frac{1}{2}\Omega_{xy}(\mathbf{k},\theta_c),  \\
	g_{yy}(\mathbf{k},\theta_c)&=\frac{1}{2}\Omega_{xy}(\mathbf{k},\theta_c),  \\
	g_{xy}(\mathbf{k},\theta_c)&=0.
\end{align}
Remarkably, the saturation of DC directly implies that the Berry curvature is (semi-)positive or (semi-)negative definite. Next, we consider a more general case in which we take $c=c_1+ic_2$, where $c_1$ and $c_2$ are real numbers. Then a general relation can be obtained, i.e.,
\begin{align}
	g_{xx}(\mathbf{k},\theta_c)&=\frac{|c|^2}{2c_2}\Omega_{xy}(\mathbf{k},\theta_c),  \\
	g_{yy}(\mathbf{k},\theta_c)&=\frac{1}{2c_2}\Omega_{xy}(\mathbf{k},\theta_c),  \\
	g_{xy}(\mathbf{k},\theta_c)&=\frac{c_1}{2c_2}\Omega_{xy}(\mathbf{k},\theta_c).
\end{align}
From Eqs. (S26), (S27), and (S28), and given that $g_{aa}(\mathbf{k},\theta_c)$ is (semi-)positive definite, it follows that the Berry curvature is (semi-)positive or (semi-)negative definite, and $c_2$ is positive (or negative) if the Berry curvature is (semi-)positive (or (semi-)negative) definite. However, $c_1$ can be either positive or negative. Thus, a saturation domain can be defined as $\mathbb{H}=\{c\in \mathbb{C}:\Im(c)>0\}$ or $\{c\in \mathbb{C}:\Im(c)<0\})$. It is similar to the definition of ideal flat band proposed in Ref. \cite{Wang21}. In their work, the corresponding parameter $w^{ab}=w^{a}w^{b*}+w^{a*}w^{b}$ can be determined by the constant null vector of the quantum geometric tensor \cite{Wang21}. Here we have generalized it to multi-band case. The saturation constant $c$ can be controlled by experimental parameters such as twisted angle, strain, lattice relaxation, strength of electron interaction, and displacement field in the moir\'e superlattice systems. If $c$ is not equal to $i$, the system can be an anisotropic FCI \cite{Claassen15} or fractional quantum Hall systems \cite{Haldane11,Yang17}. Lastly, we demonstrate that the saturation constant $c$ can be determined experimentally using the Souza-Wilkens-Martin sum rule \cite{Souza00}: $\int^{\infty}_{0}\frac{\Re\sigma_{aa}(\omega)}{\omega}d\omega=\frac{\pi e^{2}}{\hbar}\int\frac{d^2\bf{k}}{(2\pi)^2}
g_{aa}(\mathbf{k})$. To express the saturation constant in terms of the optical conductivity, we define a quantity called the normalized optical weight, $W_{aa}=\frac{\hbar}{e^2}\int^{\infty}_{0}\frac{2\Re\sigma_{aa}(\omega)}{\omega}d\omega$. By integrating both sides of Eq. (S23) and Eq. (S24) and multiplying by $1/(2\pi)$, we obtain
\begin{align}
	c_1&=\frac{\sqrt{4W_{xx}W_{yy}-\mathcal{C}}}{2W_{yy}},  \\
	c_2&=\frac{\mathcal{C}}{2W_{yy}}.
\end{align}
Here we assume that the Chern number of the occupied band, i.e., $\mathcal{C}=\int\frac{d^2\mathbf{k}}{2\pi}\Omega_{xy}(\mathbf{k},\theta_c)$, is nonzero.We can also express the Chern number using the generalized optical weight \cite{Onishi24}, i.e., $\mathcal{C}=-\frac{\hbar}{e^2}\int^{\infty}_{0}\frac{4\Im\hat{\sigma}_{xy}(\omega)}{\omega}d\omega$. Note that TC corresponds to the vortexability condition \cite{Ledwith23} and is saturated in isotropic systems, whereas DC is saturated in anisotropic systems. In the next section, we will demonstrate the relationship between the multi-band vortexability condition and the perfect circular dichroism.


\begin{figure*}[]
	\includegraphics[width=0.99\textwidth]{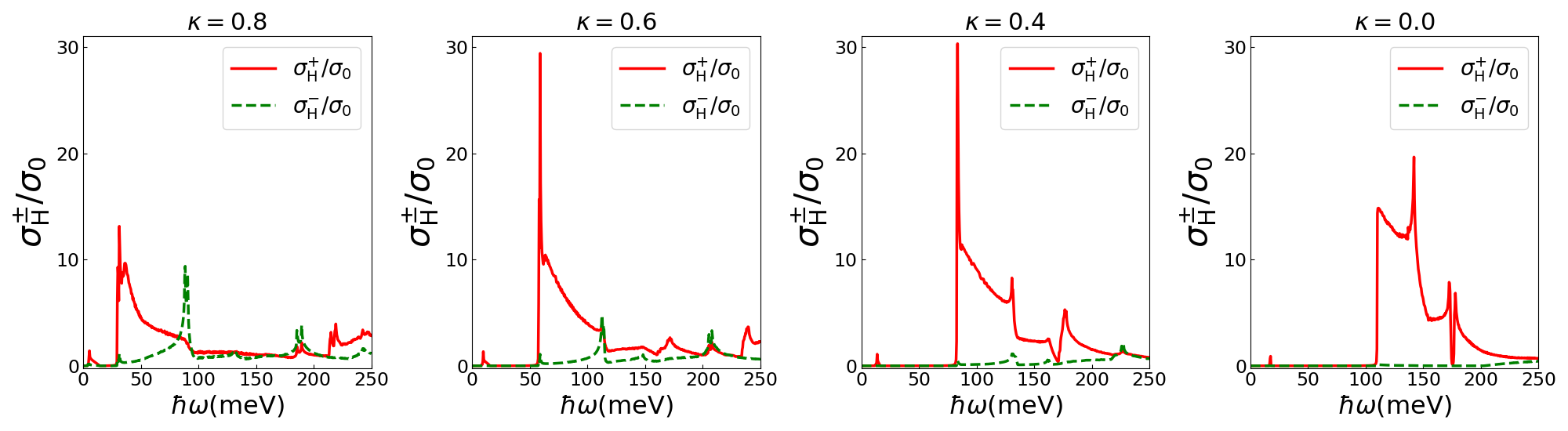}
	\caption{Evolution of the absorptive part of left-hand and right-hand circularly polarized optical conductivity as $\kappa$ varies, in units of $\sigma_{0}=\frac{e^2}{h}$. As $\kappa$ decreases, the absorptive part of left-hand circularly polarized optical conductivity, $\sigma^{-}_\text{H}(\omega)$, indicated by the green dashed line, rapidly vanishes at low energy, and the gap between the flat band and the dispersive band increases. Here we set $\theta=1.05^{\circ}$ and $\Delta_{b}=17$.
	}
\end{figure*}

\subsection{3: Multi-band Vortexability Condition and Circular Dichroism in TBG}
In this section, we demonstrate the relationship between the multi-band vortexability condition \cite{Dong23} and circular dichroism in TBG. For a set of occupied bands, we define the projection operator $P(\mathbf{k})=\sum_{n\in occ}| u_{n}(\mathbf{k}) \rangle \langle u_{n}(\mathbf{k}) |$ and its complement $Q(\mathbf{k})=I-P(\mathbf{k})$. The vortexability condition $zP(\mathbf{k})=P(\mathbf{k})zP(\mathbf{k})$ can be expressed as $P(\mathbf{k})z^{\dagger}Q(\mathbf{k})=0$ \cite{Ledwith23}, where $z=x+iy$. One can rewrite the vortexability condition as a norm identity \cite{Dong23}, i.e.,
\begin{align}
	0=||P(\mathbf{k})z^{\dagger}Q(\mathbf{k})||=\mathrm{Tr}[(P(\mathbf{k})z^{\dagger}Q(\mathbf{k}))^{\dagger}(P(\mathbf{k})z^{\dagger}Q(\mathbf{k}))].
\end{align}
Here we use the squared Frobenius norm. Substituting the expressions for $P(\mathbf{k})$ and $P(\mathbf{k})$ into Eq. (S28), we have
\begin{align}
	\sum_{m\in occ}\sum_{n\in unocc}\int\frac{d^2\bf{k}}{(2\pi)^2}|\langle u_m(\mathbf{k})|z^{\dagger}|u_n(\mathbf{k})\rangle|^2=0.
\end{align}
Note that each term in the above equation is non-negative. From this constraint, $\langle u_m(\mathbf{k})|z^{\dagger}|u_n(\mathbf{k})\rangle$ must equal zero. Importantly, this term is related to the term in the formula for left-handed circularly polarized conductivity. Since the circularly polarized velocity operator is given by $\upsilon^{\pm}=-i[x\pm iy,H]$ \cite{Dong23}, we have
\begin{align}
		\langle u_m(\mathbf{k})|\upsilon^{-}|u_n(\mathbf{k})\rangle  =-i(E_n(\mathbf{k})-E_m(\mathbf{k}))\langle  u_m(\mathbf{k})|z^{\dagger}|u_n(\mathbf{k})\rangle=0.
\end{align}
Substituting Eq. (S29) into the formula for the absorptive part of the circularly polarized optical conductivity \cite{Inoue62,Toy77,Moon13},
\begin{equation}
	\sigma^{\pm}_\text{H}(\omega)=-\frac{\pi e^2}{\hbar}\sum_{m\in occ}\sum_{n\in unocc}\int\frac{d^2\bf{k}}{(2\pi)^2}\frac{f_{mn}(\bf{k})}{E_{mn}(\bf{k})}|\langle u_m(\mathbf{k})|\upsilon^{\pm}|u_n(\mathbf{k})\rangle|^2\delta(\hbar\omega+E_{mn}(\mathbf{k})),
\end{equation}
the absorptive part of the left-handed circularly polarized optical conductivity \cite{Souza08} will vanish (perfect circular dichroism), i.e., $\sigma^{-}_\text{H}(\omega)=0$, if the multi-band vortexability condition is satisfied. Interestingly, the vortexability condition is not directly related to the vanishing of Fermi velocity because it is a diagonal element of the velocity operator. In Fig. S3, we can see that the isolated peak of $\sigma^{+}_\text{H}(\omega)$ close to zero energy is very small (see the red line), which implies that the contribution of the flat band transition to $\sigma^{+}_\text{H}(\omega)$ is also very small. In the same energy region, $\sigma^{-}_\text{H}(\omega)$ is almost zero (see the green dashed line), which implies that TC is nearly saturated in the low-energy region. As we approach the chiral limit, the zero-value region of 
$\sigma^{-}_\text{H}(\omega)$ enlarges. If all bands become flat, then $\sigma^{-}_\text{H}(\omega)$ will be completely zero. In other words, the multi-band vortexability condition is satisfied.

\bibliography{ref_CD}

\begin{thebibliography}{84}
\bibitem [{sup({\natexlab{a}})}]{supp00}%
  \BibitemOpen
  \href@noop {} {} ({\natexlab{a}}),\ \bibinfo {note} {inspired by the trace
  condition, we define isotropic quantum geometry by two conditions:
  $g_{xx}=g_{yy}$ and $g_{xy}=0$. When these conditions are violated, we refer
  to it as anisotropic quantum geometry.}\BibitemShut {Stop}%
\bibitem [{sup({\natexlab{b}})}]{supp0}%
  \BibitemOpen
  \href@noop {} {} ({\natexlab{b}}),\ \bibinfo {note} {using the components of
  the quantum metric and Berry curvature, the single-band and multiband quantum
  geometric tensors are given by $Q^{n}_{ab}(\bm{k})=\sum_{m\ne
  n}Q^{mn}_{ab}(\bm{k})=g^{n}_{ab}(\bm{k})-\frac{i}{2}\Omega^{n}_{ab}(\bm{k})$
  and $Q_{ab}(\bm{k})=\sum_{m\in occ}\sum_{n\in
  unocc}Q^{mn}_{ab}(\bm{k})=g_{ab}(\bm{k})-\frac{i}{2}\Omega_{ab}(\bm{k})$,
  respectively.}\BibitemShut {Stop}%
\bibitem [{sup({\natexlab{c}})}]{supp1}%
  \BibitemOpen
  \href@noop {} {} ({\natexlab{c}}),\ \bibinfo {note} {to reveal the
  sign-changing process of the Berry curvature toward the ideal flat band
  limit, we only consider the Berry curvature instead of the maximal Berry
  curvature in the optical bound inequality.}\BibitemShut {Stop}%
\bibitem [{sup({\natexlab{e}})}]{supp2}%
  \BibitemOpen
  \href@noop {} {} ({\natexlab{e}}),\ \bibinfo {note} {in this work, we assume
  that $c$ is a constant. Theoretically, the saturation constant $c$ can be
  adjusted by the tunable anisotropic parameters, such as the anisotropic mass
  matrix in the kinetic term \cite{Haldane11, BYang12} and the anisotropic
  interaction term \cite{Haldane11, BYang17}. Remarkably, the domain of the
  anisotropic mass matrix or lattice deformation \cite{Avron95, Levay95,
  Levay97, Tokatly07} is also the upper half of the complex plane. As a result,
  the anisotropic mass parameter and the saturation constant are related by a
  modular transformation. For more complex systems, such as those with defects,
  impurities, and disorders, $c$ can be momentum-dependent or exhibit certain
  variations across the (moiré) Brillouin zone.}\BibitemShut {Stop}%
\bibitem [{sup({\natexlab{f}})}]{supp3}%
  \BibitemOpen
  \href@noop {} {} ({\natexlab{f}}),\ \bibinfo {note} {when we set
  $c=\frac{1}{a-ib}$, we can obtain the same matrix of the almost complex
  structure.}\BibitemShut {Stop}%
\end{thebibliography}

\end{document}